\documentclass[%
 reprint,
 amsmath,amssymb,
 aps,
 prl,
]{revtex4-2}

\usepackage{graphicx}
\usepackage{dcolumn}
\usepackage{bm}
\usepackage{xcolor}
\usepackage{comment}

\begin{document}

\preprint{APS/123-QED}

\title{Reynolds-number dependence of streamwise velocity variance in wall-bounded turbulent flows}
\thanks{ctong@clemson.edu}%

\author{Chenning Tong}
 \affiliation{Department of mechanical Engineering, Clemson University, Clemson, SC 29631} 




\date{\today}

\begin{abstract}
  We propose a model for the streamwise velocity variance in wall-bounded turbulent flows. It hypothesizes that the  wall-parallel motions of the
  attached eddies induce internal turbulent boundary layers. 
  A logarithmic variance profile is obtained. 
  The peak value of the variance scaled using the friction velocity has a logarithmic dependence on the ratio the wall-normal length of the flow to
  the thickness of the internal boundary layer induced by the largest attached eddies ($\delta_o$), the latter having a  dependence on the friction Reynolds number
  in the form of a Lambert W function.
  Both the peak and the length ratio are unbounded at asymptotically large Reynolds numbers. 
  The model also predicts that the streamwise velocity fluctuations induced by the attached eddies near the
  viscous layer scale with the friction velocity; therefore  the scaled velocity variance there remains finite at asymptotically large Reynolds numbers.

\end{abstract}

\maketitle



\noindent \textit{Introduction } 

The variances of wall-parallel fluctuating velocities in the near-wall region in wall-bounded turbulent flows are important statistic for fundamental
understanding of such flows and for engineering applications.
For example, in the zero-pressure-gradient turbulent boundary layer, the streamwise derivative of the streamwise velocity variance, $\partial \overline{u_1^2}/\partial x$,
is one of the quantities determining the higher-order mean velocity profile. Also, dispersion in such flow depends on the spanwise velocity variance. 
Townsend \cite{Townsend76} predicted using the attached-eddy model that the peak of the scaled variance $\overline{u_1^{2+}}=\overline{u_1^2}/u_*^2\sim \ln Re_*=\ln u_*\delta/\nu$,
where $u_*$, $Re_*$, $\nu$, and $\delta$
are the friction velocity, the friction Reynolds number, the kinematic viscosity, and the wall-normal scale of the flow, which can be the half channel width, pipe radius, or
the boundary layer thickness. This prediction is in apparent contradiction of the law-of-the-wall \cite{Prandtl1925} type scaling which requires
that the scaled variance is independent of $Re_*$.
In recent years, there have been many efforts devoted to this issue
\cite{Marusic2003,Morrison2004,Hoyas2006,Marusic2010b,Smits2011,Mathis2011,Monkewits2015,Marusic2017,Marusic2019,Chen2021,Chen2022}. 

Experimental and direct numerical simulation results have both shown that $\overline{u_1^{2+}}$ has a (inner) peak near $y^+=u_*y/\nu=15$ and the peak value appears
to increase with $Re_*$ \cite{Hoyas2006,Marusic2015,Samie2018},
consistent with Townsend's prediction, where $y$ is the wall-normal coordinate. On the other hand, Chen \& Sreenivasan \cite{Chen2021} argued that this prediction will lead to
an unbounded viscous-scaled dissipation rate as $Re_* \rightarrow \infty$, which is inconsistent with the maximum of  $1/4$ for the scaled production.
They proposed a model based on
the dissipation rate, which predicts that the inner peak of $\overline{u_1^{2+}}$ will asymptote to a value of approximately 12. They also argue that
their prediction fits the existing data better than Townsend's prediction.

While the asymptotic behavior of the inner peak has not been fully resolved, experimental results \cite{Hultmark2012,Hultmark2013,Vallikivi2015} have shown that
another peak is emerging further away from the wall,
with its location in terms of $y^+$ increasing with $Re_*$. This peak, often referred to as the ``outer'' peak, has has also received some attention in recent years.
Pullin \cite{Pullin2013} obtained a logarithmic dependence of the peak value of
$\overline{u_1^{2+}}$ on $Re_*$.
In this work, we propose a model for the streamwise velocity variance by hypothesizing the existence of internal turbulent boundary layers induced by the  attached eddies. 
The model provides both the Reynolds number dependence  of the ``outer'' peak and the contributions from attached eddies to the velocity variance near the viscous layer. The latter
will also help address the issue of boundedness of the inner peak near the viscous layer.

\vspace{6pt}

\noindent \textit{Internal boundary layer analysis }

In the traditional understanding of the attached eddy model and the interpretations of experimental and simulation results based on the understanding,
the wall-parallel motions of the attached eddies are considered to extend towards the wall until viscous effects damp them (at $y^+\sim 15$) (e.g.,\cite{Townsend76}).
In this work we argue that because these near-wall motions are approximately two-dimensional, the nonlinear interactions among them are weak.  They have significant interactions
only with scales of order $y$ (or smaller), thereby generating their own, or internal, turbulent boundary layers with a ``free-stream'' velocity of order $u_*$;
therefore, the  magnitudes of the  velocities of these motions will begin to  decrease well before
the viscous effects on them are dominant. By analyzing the boundary layers, the contributions of the attached eddies to the velocity variance can be calculated and the peak
value of $\overline{u_1^{2+}}$ and its location can be estimated. Note that we do not consider the streamwise fluctuations due to mean shear production in the
near-wall layer considered, which are not part of the contributions of the attached eddies, and has been investigated
by \cite{Chen2021}.

 For convenience we analyze the properties of the internal boundary layers in horizontal Fourier space. 
For a horizontally homogeneous velocity field, $u_i$,
we define its horizontal Fourier transform over a horizontal physical domain of size $\mathcal{L} \times \mathcal{L}$ as \cite{TD19}
\begin{equation}
\hat{u}_i(\bm{k}) = \frac{1}{\mathcal{L}^2}\int_{0}^{\mathcal{L}}u_i(\bm{x},t)e^{-i\bm{k}\bm{x}}d\bm{x},
\end{equation}
where $\bm{k}$ is the horizontal wavenumber vector.
Dividing the usual definition of the transform by $\mathcal{L}^2$ gives the Fourier transform the same dimension as well as the same scaling properties as the velocity,
which is convenient for the scaling analysis.

We first consider the internal turbulent boundary layer induced by the largest attached eddies with a scale of $\delta$.
 The equation for $ \hat{u}_{1}(\bm{k})$ can be written as \cite{TD19}
\begin{equation}
\begin{aligned}
&\frac{\partial \hat{u}_{1}(\bm{k})}{\partial t} + \mathcal{L}^2\int\hat{u}_{3}(\bm{k'})\frac{\partial{\hat{u}_{1}(\bm{k-k'})}}{\partial{x_{3}}}d\bm{k}' + \mathcal{L}^2\int\hat{u}_{1}ik_{1}'\hat{u}_{1}d\bm{k}' \\ & = -ik_{1}\hat{p}+\nu \frac{\partial^2 \hat{u}_{1}(\bm{k})}{\partial x_3^2}.\\
\end{aligned}
\label{momentum}
\end{equation}
In the inner layer of the internal boundary layer, the viscous stress divergence is important.
For $k\sim 1/\delta_i$, the dominant terms are the production term (the second on term on the LHS in (\ref{momentum})) and the viscous term, where $\delta_i$ is the viscous length
of the internal boundary layer.
The production term has the contributions mostly from $k' \sim 1/\delta_i$ and
$|\bm{k}-\bm{k'}|\sim 1/\delta$, with the latter playing the role of the ``mean'' velocity. These terms scale as 
\begin{equation}
 u_* \frac{\hat{u}_{1}(1/\delta)}{\delta_i} \sim \nu \frac{\hat{u}_{1}(1/\delta_i)}{\delta_i^2}.
  \label{inner_di}
\end{equation}
Here  the fluctuations of the attached eddies $\hat{u}_{1}(\bm{k-k'})$ interact with
the wall-normal fluctuations of scale $\delta_i$, $\hat{u}_{3}(\bm{k'})$, which are due to the usual mean-shear production/pressure-strain-rate interaction and
scale with $u_*$, and are part of the ``background'' turbulence in which  the internal boundary layer develops.

For $k\sim 1/\delta$ in the inner layer, the dominant terms in (\ref{momentum}) are the induced shear stress term (the second on term on the LHS in (\ref{momentum})) and the viscous term.
The former has the contributions mostly from $k' \sim |\bm{k}-\bm{k'}|\sim 1/\delta_i$, and
has the same scaling as the viscous term
\begin{equation}
  u_* \frac{\hat{u}_{1}(1/\delta_i)}{\delta_i} \sim \nu \frac{\hat{u}_{1}(1/\delta)}{\delta_i^2}.
  \label{inner_d}
\end{equation}
Here the viscous term is the ``mean'' viscous stress derivative.
From the scaling in  (\ref{inner_di}) and (\ref{inner_d}), we obtain $\hat{u}_{1}(1/\delta_i) \sim \hat{u}_{1}(1/\delta)$, which we denote as $v_*$, and $\delta_i \sim \nu/u_*=\delta_\nu$.
Therefore, the induced shear stress scales as $u_*v_*$.
The ``mean'' velocity derivative in the inner layer has the form
\begin{equation}
  \frac{\partial \hat{u}_{1}(1/\delta)}{\partial y} = \frac{v_*}{\delta_\nu}f(\frac{y}{\delta_\nu}).
  \label{inner_derivaive}
\end{equation}
Note that the internal boundary layer is a stochastic process; therefore, (\ref{inner_derivaive}) and similar relations are valid in an average sense.

In the outer layer  of the induced boundary layer,  the induced-stress and  advection terms (the second and third terms on the LHS in (\ref{momentum})) are important.
They scale as 
\begin{equation}
u_* \frac{v_*}{\delta_o}\sim \frac{u_*^2}{\delta}, 
\label{outer}
\end{equation}
where $\delta_o$ is the thickness of the internal boundary layer. Note again that the ``free-stream'' velocity  of the internal boundary layer scales as $u_*$.
In the outer layer $u_1$ is dominated by $k\sim 1/\delta$. The induced-stress term is dominated by $k' \sim |\bm{k}-\bm{k'}|\sim 1/\delta_o$, whereas
the advection term (the third term on the LHS in (\ref{momentum})) is dominated by contributions from $k' \sim |\bm{k}-\bm{k'}|\sim 1/\delta$.

From (\ref{outer}) we obtain $u_*/v_*\sim \delta/\delta_o$.
The ``mean'' velocity derivative in the outer layer has the form
\begin{equation}
  \frac{\partial \hat{u}_{1}(1/\delta)}{\partial y} = \frac{v_*}{\delta_o}F(\frac{y}{\delta_o}).
  \label{outer_derivaive}
\end{equation}

Asymptotically matching (\ref{inner_derivaive}) and (\ref{outer_derivaive}) we obtain a log law and a logarithmic friction law,
\begin{equation}\label{log_law}
  \frac{\hat{u}_{1}(1/\delta)}{v_*}=\frac{1}{\kappa}\ln\frac{ y}{\delta_\nu} +B, \ \  \frac{\hat{u}_{1}(1/\delta)-u_*}{v_*}=\frac{1}{\kappa}\ln\frac{ y}{\delta_o} +C,
\end{equation}
\begin{equation}\label{friction_law}
  \frac{u_*}{v_*}=\frac{1}{\kappa}\ln\frac{\delta_o}{\delta_\nu} +B-C=\alpha \frac{\delta}{\delta_o},
\end{equation}
where $\alpha$  is a non-dimensional coefficient of order one. Again, these relations are valid  in an average sense.
A similar analysis can also be carried out for smaller attached  eddies with $k> 1/\delta$. For such eddies  $\delta_o(k)$ and $v_*(k)$ are functions of $k$.

\vspace{6pt}

\noindent {\it Velocity variance - outer peak}

Based on the above analysis, it is clear from (\ref{friction_law}) that
$\delta_o(k)< \delta_o(1/\delta)$. Therefore, the contribution to the velocity variance from these eddies is $u_*^2\sim \phi_u(k)k$ for $1/k>y>\delta_o(1/\delta)$,
where $\phi_u(k)$ is the spectrum of $u_1$.
Integrating $\phi_u(k)$ from $1/\delta$ to $1/y$ results in the variance
\begin{equation}\label{log_variance}
 \overline{u_1^{2+}}= \frac{\overline{u_1^2}}{u_*^2}=A_v\ln\frac{\delta}{y}+B_v, 
\end{equation}
the same as given in \cite{Townsend},
where $A_v$ and $B_v$ are non-dimensional coefficients. Because $\hat{u}_{1}(1/\delta)$ decreases with $y$ for $y< \delta_o(1/\delta)$, the variance will peak at $y_p\sim \delta_o(1/\delta)$,
with a peak value of
\begin{equation}\label{peak}
  \overline{u_{1p}^{2+}}=A_v\ln\dfrac{\delta}{\delta_o}+B'_v,
  \end{equation}
where $B'_v$ is another non-dimensional constant. Equation (\ref{friction_law}) can be rewritten as
\begin{equation}\label{friction_Re}
  \frac{u_*}{v_*}=\alpha\frac{\delta}{\delta_o}=\frac{1}{\kappa}(\ln\frac{\delta_o}{\delta}+\ln Re_*) +D, 
\end{equation}
which can be further written as
\begin{equation}\label{Re}
\frac{\delta}{\delta_o}e^{\alpha\kappa\delta/\delta_o}=e^{\kappa D}Re_*.
\end{equation}
We set $\alpha=1$ as it can be absorbed into $\delta/\delta_o$ and $D$ by redefine $\delta/\delta_o$.
We recognize $\delta/\delta_o$ as the Lambert $W$ function (with $e^{\kappa D}Re_*$ as the independent variable).
The dependence of $\overline{u_{1p}^{2+}}$ on $Re_*$ can be obtained from (\ref{peak}) and (\ref{friction_Re}) as
\begin{equation}\label{peak1}
  \overline{u_{1p}^{2+}}=A_v\ln W(e^{\kappa D}Re_*)+B'_v,
  \end{equation}
and shows that the ratio $\delta/\delta_o$ and the peak value $\overline{u_{1p}^{2+}}$ grows unbounded as $Re_* \rightarrow \infty$.
The unboundedness of $\delta/\delta_o$ indicates that the peak location the peak occurs deeper into the log layer as $Re_*$ increases. On the other hand,
since $\delta/\delta_o$ increases slower than $Re_*=\delta/\delta_\nu$, the viscous-scaled peak location $y^+_p\sim \delta_o/\delta_\nu \rightarrow \infty$ as
$Re_* \rightarrow \infty$, i.e., $y^+_p$ moves further away from the viscous layer as $Re_*$ increases.

\begin{figure}
\centering
\includegraphics[width=3.6in,height=2.4in]{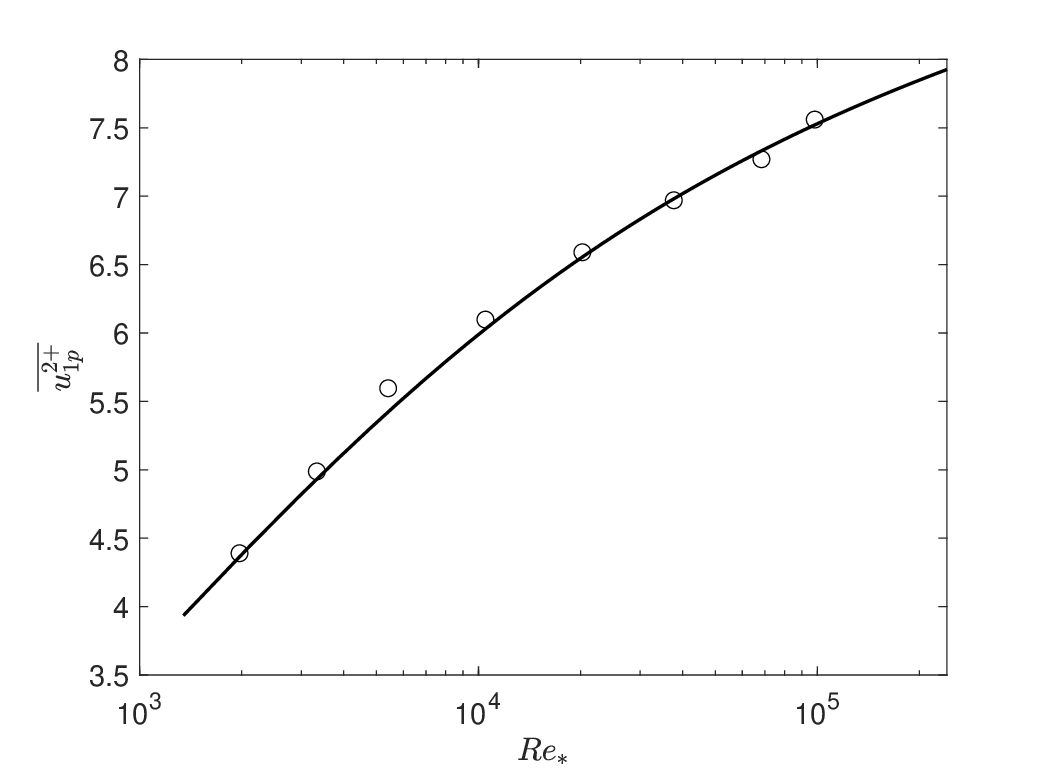}
\caption{Outer peak value of non-dimensional velocity variance as a function of $Re_*$. Circles: experimental data of pipe flows 
shown in \cite{Vallikivi2015}; Solid line: Theoretical prediction of equation (\ref{peak1}).}
\vspace{0.0in}
\label{fig1}
\end{figure}

We now use experimental data of \cite{Hultmark2013} to evaluate the model. Using the values of $\overline{u_{1p}^{2+}}$  at two different $Re_*$, from (\ref{peak}) we can obtain
\begin{equation}\label{}
  \overline{u_{1p}^{2+}}_2- \overline{u_{1p}^{2+}}_1=A_v\ln\dfrac{(\delta/\delta_o)_2}{(\delta/\delta_o)_1},
  \end{equation}
where $A_v=1.24$ for the data (\cite{Hultmark2013}).
From (\ref{friction_Re}) we obtain
\begin{equation}\label{}
(\frac{\delta}{\delta_o})_2-(\frac{\delta}{\delta_o})_1=-\frac{1}{\kappa}\ln\dfrac{(\delta/\delta_o)_2}{(\delta/\delta_o)_1}+\frac{1}{\kappa}\ln \frac{Re_{*2}}{Re_{*1}}, 
\end{equation}
From these equations we obtain the values of $(\dfrac{\delta}{\delta_o})_1$ and $(\dfrac{\delta}{\delta_o})_2$.
Then from (\ref{peak}) and (\ref{friction_Re}) we obtain the values $B'_v=5.9$ and $D=-21.4$ respectively.
Figure \ref{fig1} shows the dependence of $\overline{u_{1p}^{2+}}$ on $Re_*$ given by (\ref{peak1}). The model describes the experimental data well for the $Re_*$ range $\approx 2000-100,000$,
supporting the model hypothesis that the attached eddies induce internal turbulent boundary layers to affect the behaviors of $\overline{u_{1}^{2+}}$.

The $Re_*$ dependence of $\overline{u_{1p}^{2+}}$ highlights the limitations of the law of the wall, which was proposed for the mean velocity and is a mean field theory.
The velocity variances are fluctuation statistics. In general, there is no reason for them to follow a mean field theory, even in wall flows with a single
velocity scale ($u_*$). Failures of mean field theories when applied to fluctuations occur more often in flows with multiple velocity scales, such as in the atmospheric
boundary layer, where two velocity scales are present, one due to mean shear ($u_*$) and
the other due to buoyancy. The Monin-Obukhov similarity theory \cite{MO54}, developed as a mean field theory, is successful in scaling the mean velocity in the surface layer
of the atmospheric boundary layer, but
fails to scale the wall-parallel velocity variances and spectra.
Since fluctuating velocities can have contributions from a wide range of scales, to account for these contributions, a multipoint (minimum of two-point) theory
is needed. \cite{TN15,TD19} developed the multipoint Monin-Obukhov similarity theory, successfully overcoming the limitations of the original theory.
The analysis of the streamwise velocity fluctuations in Fourier space (Eq.~\ref{momentum}) in the present work is essentially a two-point theory, and therefore is capable of
successfully explaining the outer peak of the streamwise velocity variance.

\vspace{6pt}
 
\noindent {\it Velocity variance near viscous layer}

Since its outer peak location  scales as $y_p\sim \delta_o$, the total contribution to $\overline{u_{1}^{2+}}$ from the attached eddies begins to decrease when moving closer towards the wall.
We now estimate $\overline{u_{1}^{2+}}$ for $y\sim \delta_\nu$. In this layer, the horizontal velocity scale for the attached eddies of scale $\delta(k)$ is
$v_*(k)$. For the attached eddies of scale $1/k<\delta$, it is given by
\begin{equation}\label{}
  {v_*(k)}=u_*\frac{\delta_o(k)}{1/k}=u_*\delta_o(k)k.
\end{equation}
The friction law is
\begin{equation}\label{friction_law_k}
  \frac{u_*}{v_*(k)}=\frac{1}{k\delta_o}=\frac{1}{\kappa}\Big\{\ln(k\delta_o(k))+\ln Re_k\Big\} +D, 
\end{equation}
where $Re_k=u_*/(k\nu)$. The contribution to $\overline{u_1^2}$ is $v_*^2(k)=\phi_u(k)k$. Therefore, $\phi_u(k)=v_*^2(k)/k$.
Integrating it from $1/\delta$ to $1/(\beta\delta_\nu)$,
where $\beta >1$ is a non-dimensional constant such that an internal boundary layer exists for attached eddies of scale $\beta \delta_\nu$ or larger, we have
\begin{equation}\label{integral}
  \overline{u_1^2} \sim \int_{1/\delta}^{1/(\beta\delta_\nu)}\frac{v_*^2(k)}{k}dk=u_*^2\int_{1/\delta}^{1/(\beta\delta_\nu)}\frac{\delta_o^2(\kappa)}{k(1/k)^2}dk.
\end{equation}
It is unclear how to proceed with direct integration. Instead we
define $W(k)=1/(k\delta_0) $. Equation (\ref{friction_law_k}) can be written as
\begin{equation}\label{friction_law_w}
  W(k)=\frac{1}{\kappa}\Big\{-\ln W(k)+\ln Re_k\Big\} +D. 
\end{equation}
 The differential of the this expression is
\begin{equation}\label{friction_law_dw}
  dW(k)=\frac{1}{\kappa}\Big\{-d\ln W(k)-d\ln k\Big\}. 
\end{equation}

The integral in (\ref{friction_law_k}) can be written as
\begin{equation}\label{}
  \int_{1/\delta}^{1/(\beta\delta_\nu)}\frac{d\ln k}{ W^2}=\int_{\delta/\delta_0(1/\delta)}^{\beta\delta_\nu/\delta_0(1/\beta\delta_\nu)}\frac{-(\kappa dW+d\ln W)}{ W^2}  \nonumber
\end{equation}
\begin{equation}\label{}
  = \frac{\kappa}{ W} \Bigg|_{\delta/\delta_0(1/\delta)}^{\beta\delta_\nu/\delta_0(1/\beta\delta_\nu)}+ \frac{1}{ W^2} \Bigg |_{\delta/\delta_0(1/\delta)}^{\beta\delta_\nu/\delta_0(1/\beta\delta_\nu)}\nonumber
\end{equation}
\begin{equation}\label{}
  = \frac{\kappa}{\dfrac{\beta\delta_\nu}{\delta_0(1/\beta\delta_\nu)}}-\frac{\kappa}{ \dfrac{\beta\delta}{\delta_o(1/\delta)}}
  +\frac{1}{\dfrac{2\beta^2\delta^2_\nu}{\delta^2_o(1/\beta\delta_\nu)}}-\frac{1}{ \dfrac{2\beta^2\delta^2}{\delta^2_o(1/\delta)}},
\end{equation}
which remains finite as $\delta/\delta_o(1/\delta) \rightarrow \infty$ ($Re_* \rightarrow \infty$).
This result indicates that the contributions to $\overline{u_1^2}$ from the attached eddies scale as $u_*^2$ for $y\sim \delta_\nu$, and do not lead to an unbounded peak near the viscous layer
at asymptotically large $Re_*$.
Our model therefore suggests that the asymptotic value of the inner peak of $\overline{u_{1}^{2+}}$ as $Re_* \rightarrow \infty$ is determined by  the
local turbulence dynamics among production, dissipation, and transport, which is argued to result in a finite inner peak value $\overline{u_{1p}^{2+}}$ because
the viscous scaled production has a maximum of $1/4$  \cite{Chen2021}.

 
\vspace{6pt}

\noindent \textit{Conclusions}

We proposed a model for the contributions from the attached eddies to the streamwise velocity variance based on the internal boundary layers induced by the attached eddies.
A spectral analysis shows that the viscous length of these boundary layers is the usual viscous length $\delta_\nu=\nu/u_*$. The momentum balance in the outer layer of each
internal boundary layer is between the advection and wall-normal shear-stress derivative. A logarithmic friction law with the ``free-stream'' velocity as $u_*$ is obtained,
which gives the ``mean'' velocity scales of the internal boundary layers. A peak in the scaled  streamwise velocity variance $\overline{u_{1}^{2+}}$ is predicted, which
is unbounded as $Re_* \rightarrow \infty$. The peak location moves deeper into the log layer as $Re_*$ increases. In the mean time it moves to larger $y^+$, further
away from the viscous layer. The model is able to explain the experimental data of pipe flows \cite{Hultmark2013} well.

The model also predicts that the total contribution from the attached eddies to the streamwise velocity variance near the viscous layer scales with the
square of the friction velocity. This result combined with the finite contributions from the local turbulence dynamics
among production, dissipation, and transport \cite{Chen2021} suggests that that the peak near $y^+=15$ is  finite as $Re_* \rightarrow \infty$.

The results in the present study also have implications for the second- and higher-order mean velocity profile in turbulent boundary layers.

\begin{acknowledgments}
This work was supported by the National Science Foundation under grant AGS-2054983.
\end{acknowledgments}

\vspace{-12pt}

\bibliography{../clemson1}
\end{document}